\documentclass[runningheads]{llncs}
\usepackage{graphicx}
\usepackage{booktabs} % For formal tables
\usepackage{graphicx}
\usepackage{subcaption}
\usepackage{soul}
\usepackage{algorithm}
\usepackage{algorithmic}
\usepackage{multirow}
\usepackage{array}
\usepackage{cite}
\usepackage{amsmath}
\usepackage[font=small,skip=0.001cm]{caption}
\usepackage{tabularx}
\usepackage{float}
\begin{document}
\title{A Blockchain-Enhanced Framework for Privacy and Data Integrity in Crowdsourced Drone Services}
\titlerunning{Blockchain-Enhanced Framework for Privacy and Data Integrity}

\author{Junaid Akram\inst{1} \and  Ali Anaissi\inst{1,2} }
\authorrunning{J. Akram et al.}
\institute{The University of Sydney, Camperdown, NSW 2008, Australia
\email{jakr7229@uni.sydney.edu.au, ali.anaissi@sydney.edu.au} \and
University of Technology Sydney, Ultimo, NSW 2007, Australia
}
\maketitle              % typeset the header of the contribution
\begin{abstract}
We present an innovative framework that integrates consumer-grade drones into bushfire management, addressing both service improvement and data privacy concerns under Australia's Privacy Act 1988. This system establishes a marketplace where bushfire management authorities, as data consumers, access critical information from drone operators, who serve as data providers. The framework employs local differential privacy to safeguard the privacy of data providers from all system entities, ensuring compliance with privacy standards. Additionally, a blockchain-based solution facilitates fair data and fee exchanges while maintaining immutable records for enhanced accountability. Validated through a proof-of-concept implementation, the framework's scalability and adaptability make it well-suited for large-scale, real-world applications in bushfire management.

\end{abstract}

\section{Introduction}
The advent of consumer-grade drones has notably influenced several sectors, particularly in environmental management and emergency services \cite{akram2024ddrm,akram2022bc}. These technologies have revolutionized bushfire management, enabling residents in fire-prone areas to actively participate in detection and management efforts through crowdsourced drone services \cite{akram2024d2xchain}. Drones facilitate the collection of real-time, comprehensive data, significantly enhancing traditional bushfire management systems \cite{khan2021spice}. However, this advancement introduces new challenges, especially concerning privacy and data protection \cite{akram2024dronessl}. As drone operators collect sensitive data, it becomes essential to address privacy concerns and ensure compliance with stringent legal frameworks, such as Australia's Privacy Act 1988.

To mitigate these challenges, our study proposes a privacy-preserving framework tailored to the specific needs of crowdsourced drone services in bushfire management. By implementing local differential privacy, our architecture allows drone operators to anonymize their data independently before sharing it with data consumers, such as bushfire management authorities \cite{10547221}. This approach protects the privacy of data providers while ensuring compliance with legal standards \cite{10492460}. Furthermore, the framework introduces a system operator responsible for aggregating data and applying filtering rules. This additional layer of oversight prevents any misuse of sensitive information by any party involved, including system operators themselves \cite{akram2024priv}.

Our system establishes a unique marketplace where bushfire management authorities, as data consumers, acquire critical information from drone operators, who serve as data providers. These operators gather key data that is instrumental in generating essential statistics for effective bushfire management. The implementation of local differential privacy forms the foundation of our system, ensuring robust protection of data providers' privacy from all entities, including the system operator. Additionally, the integration of blockchain technology facilitates fair data and fee exchanges, while maintaining immutable records for increased accountability and transparency. The scalability of this design has been validated through a proof-of-concept implementation, underscoring its applicability to large-scale data collection scenarios and its practicality in real-world bushfire management settings.

The major contributions of our work are as follows:

\begin{enumerate} 

\item We propose a framework leveraging local differential privacy for secure crowdsourced drone data handling, ensuring provider privacy and compliance with laws like Australia’s Privacy Act 1988. 

\item Our blockchain integration creates an immutable operations record, enhancing security and transparency in data exchanges. 

\item We introduce a consent mechanism promoting transparency and voluntary participation, crucial for public trust and ethical data sharing. 

\item The paper conducts a risk analysis, presenting mitigation strategies for privacy and operational challenges in crowdsourced drone services. 

\item Empirical evaluation through proof-of-concept validates the framework’s effectiveness and adaptability for large-scale scenarios, particularly in bushfire management. 

\end{enumerate}

\section{Related Work} \label{related_work}

Differential privacy, traditionally implemented in centralized systems where trusted entities manage user data and apply noise for privacy, has been widely used across domains such as recommendation systems and intelligent transportation \cite{  friedman2014privacy,10535995}. In contrast, our method enhances privacy directly at the data source through local differential privacy, albeit requiring additional data to achieve accuracy. This approach separates statistical calculations from filtering entities, addressing scenarios where filtering entities, such as university ranking authorities, should not access statistical outcomes. Blockchain technology has been explored in recent studies for data exchange, as seen in Dimitriou and Mohammed’s use of Bitcoin for smart-grid data, though privacy risks persist post-payment \cite{dimitriou2020fair}. Duan et al.'s use of encrypted data in smart contracts is another example, yet our system circumvents the need for blockchain data storage, favoring scalability for large user scenarios \cite{duan2019aggregating}. While cryptographic methods like secure multi-party computation offer accuracy, they heighten privacy risks and computational demands \cite{burkhart2010sepia}. Our framework balances privacy and accuracy, applying differential privacy mechanisms locally, which contrasts with centralized systems like Gai et al.'s Industrial IoT model, where a central "Optimization Server" aggregates data \cite{gai2019differential}. By eliminating inter-drone communication and applying privacy directly at the source, our approach offers an innovative solution for decentralized environments that prioritizes privacy while maintaining data accuracy.

\section{System Model} \label{system_model}

\begin{figure}[t]
    \centering
    \includegraphics[width=0.6\columnwidth]{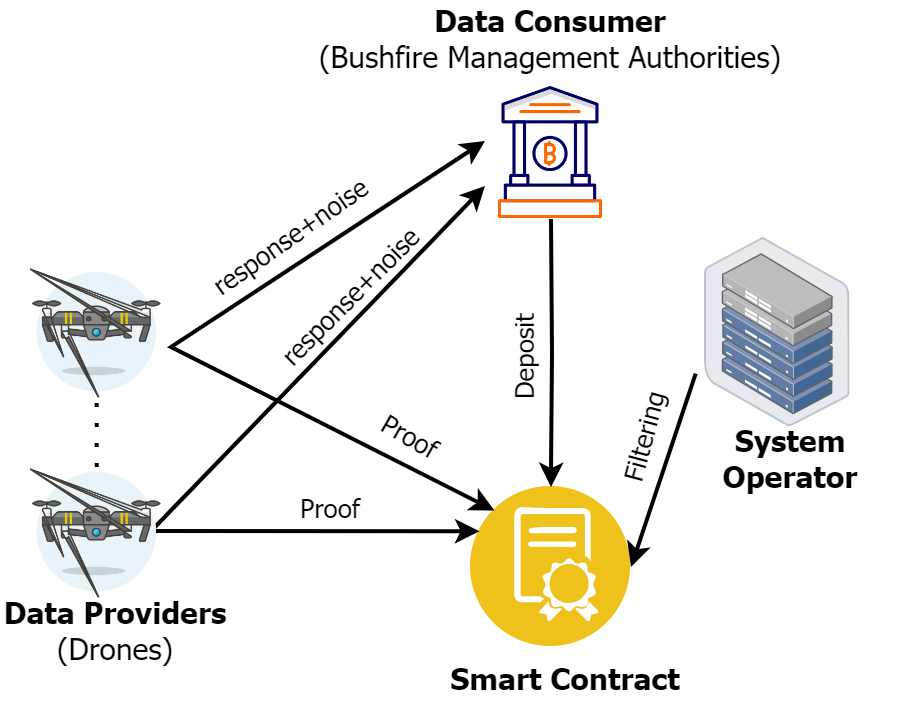}
    \caption{System Overview.}
    \label{fig:figure2}
\end{figure}

As illustrated in Fig. \ref{fig:figure2}, the system comprises three entities: bushfire management authorities, drone operators, and the system operator \cite{FOTIOU2021100022}. The bushfire management authority formulates a query \( Q = \{c_1, c_2, \ldots, c_n\} \), where \( c_1, c_2, \ldots, c_n \) represent data points or questions relevant to bushfire management. Drone operators, equipped with consumer-grade drones, collect the required data and apply local differential privacy before submitting their responses. The system operator manages the response aggregation and applies filtering rules based on the requirements specified by the bushfire authority. The response \( R \) from each operator is represented as a bit vector, where the randomized response method determines \( r_i \) for each \( c_i \) in the query \( Q \). Local differential privacy ensures the privacy of each drone operator's exact response, even from the system operator.

The process involves the bushfire authority paying only after receiving the specified number of responses \( NR \), while the system operator is responsible for distributing compensation to the drone operators. This ensures a fair exchange, where responses remain private and anonymous until the agreed number of responses is fulfilled. The system leverages blockchain, specifically Ethereum smart contracts, which store cryptographic hashes \( H(C_R) \) of the responses to guarantee immutability and transparency of the transactions. The privacy-preserving trust relationships within the system ensure that neither the system operator nor the bushfire authority has access to the individual data responses. Additionally, the filtering process is publicly verifiable without compromising the privacy of the statistical results.

The system architecture relies on a secure cryptographic protocol. A symmetric encryption key \( sk \) is generated from the nonces \( n_1 \) and \( n_2 \) using HMAC, which is used to encrypt the responses from the drone operators. The encrypted responses \( C_R \) are submitted through Ethereum addresses known to the system operator. The system operator verifies the hash \( H(C_R) \) and records it on the blockchain. The response filtering is conducted dynamically based on a bit vector \( F \), where the bit \( f_i \) is set to 1 if the response satisfies the filtering criteria, or 0 otherwise. The final set of filtered responses is published along with a hash \( H(F) \) to ensure transparency. Once the required number of valid responses \( N_R \) is reached, the system operator finalizes the process, ensuring fairness, privacy, and scalability.

\section{Design and Implementation} \label{sec_design}

Our system establishes a secure communication protocol between drone operators and the system operator, using cryptographic primitives. Each drone operator has a pre-shared secret key \( psk \), a hash function \( H(\text{data}) \), and a symmetric encryption method \( E(\text{key}, \text{data}) \). The drone operators submit their responses through an Ethereum address, ensuring that their identities remain confidential. The query, created by the bushfire authority, can be represented as a vector \( Q = \{c_1, c_2, \ldots, c_n\} \), with each \( c_i \) representing a specific data point. The drone operators respond using local differential privacy, ensuring their data remains private from the system operator and other entities.

To ensure immutability and fairness, the system relies on a smart contract that stores the cryptographic hash of the query and the responses. This contract enforces the rules for payment distribution and records the status of the responses. A blockchain-based approach ensures that the costs associated with the smart contract are independent of the number of responses, which makes the system scalable for large-scale applications like bushfire management. The responses are processed in the form of a bit vector \( R \), where each element \( r_i \) corresponds to a randomized response for the choice \( c_i \) in the query \( Q \). The encrypted response vector \( C_R = E(sk, R) \) is submitted, and its hash \( H(C_R) \) is stored in the smart contract along with the Ethereum address of the drone operator.

Once the responses are submitted, the system operator filters the responses using a bit vector \( F \), which dynamically adjusts based on the criteria set by the bushfire authority. The bit \( f_i \) in \( F \) is set to 1 if the response meets the filtering requirements, and 0 otherwise. When the number of valid responses reaches \( N_R \), the system operator completes the filtering process and publishes the final hash \( H(F) \) in the smart contract. The drone operators are then compensated according to the agreement, and the bushfire authority gains access to the encrypted responses \( C_R \), which they can decrypt using the key \( sk \).

The payment process is governed by a fair exchange protocol implemented in the smart contract. The bushfire authority deposits the agreed amount into the smart contract upon receiving \( N_R \) valid responses. The system operator then reveals the second nonce \( s_2 \), which allows the contract to verify the integrity of the transaction by checking if \( H(s_2) \) matches the stored hash. Once verified, the smart contract transfers the deposit to the system operator, who distributes the payments to the drone operators. The bushfire authority can then decrypt the responses using the derived key \( sk = H(s_1 || s_2) \) and extract the necessary statistics.

The final step in the process is for the bushfire authority to decrypt the noisy responses and derive the statistics for bushfire management. The statistical data collected from the drone operators provides a probabilistic representation of the state of the environment, which is useful for early detection and management of bushfires. The use of local differential privacy ensures that the responses are anonymized, while the blockchain-based architecture guarantees the integrity and transparency of the data collection process. This combination of technologies creates a scalable, privacy-preserving framework suitable for large-scale data collection scenarios in bushfire management.

\section{Analysis and Discussion} \label{analysis}

\subsection{Local Differential Privacy Efficiency}
\begin{figure}[t]
    \centering
    \begin{subfigure}[b]{0.24\textwidth}
        \centering
        \includegraphics[width=\textwidth]{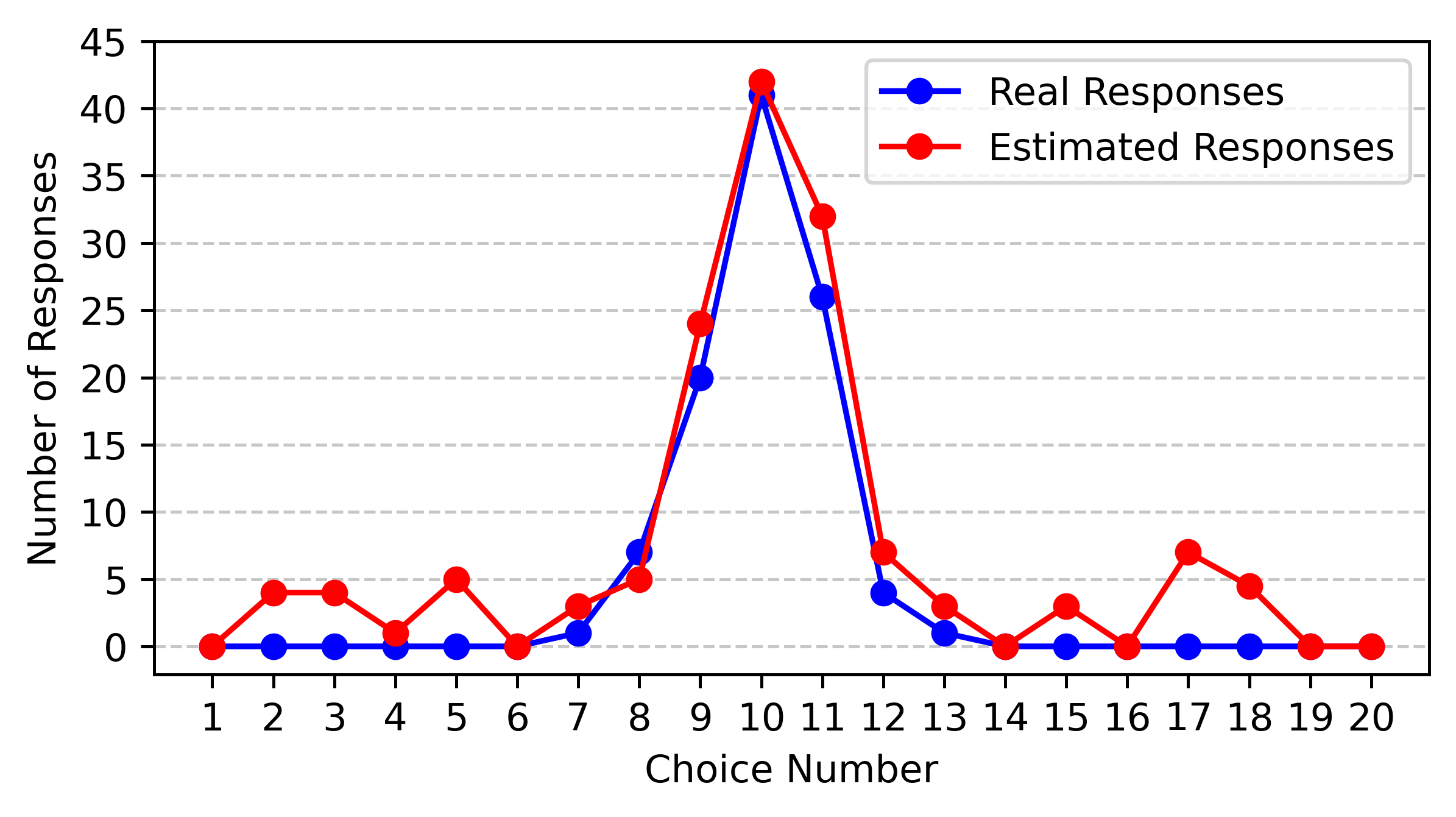}
        \caption{500 Drones}
        \label{fig:subfig1}
    \end{subfigure}
    \hfill
    \begin{subfigure}[b]{0.24\textwidth}
        \centering
        \includegraphics[width=\textwidth]{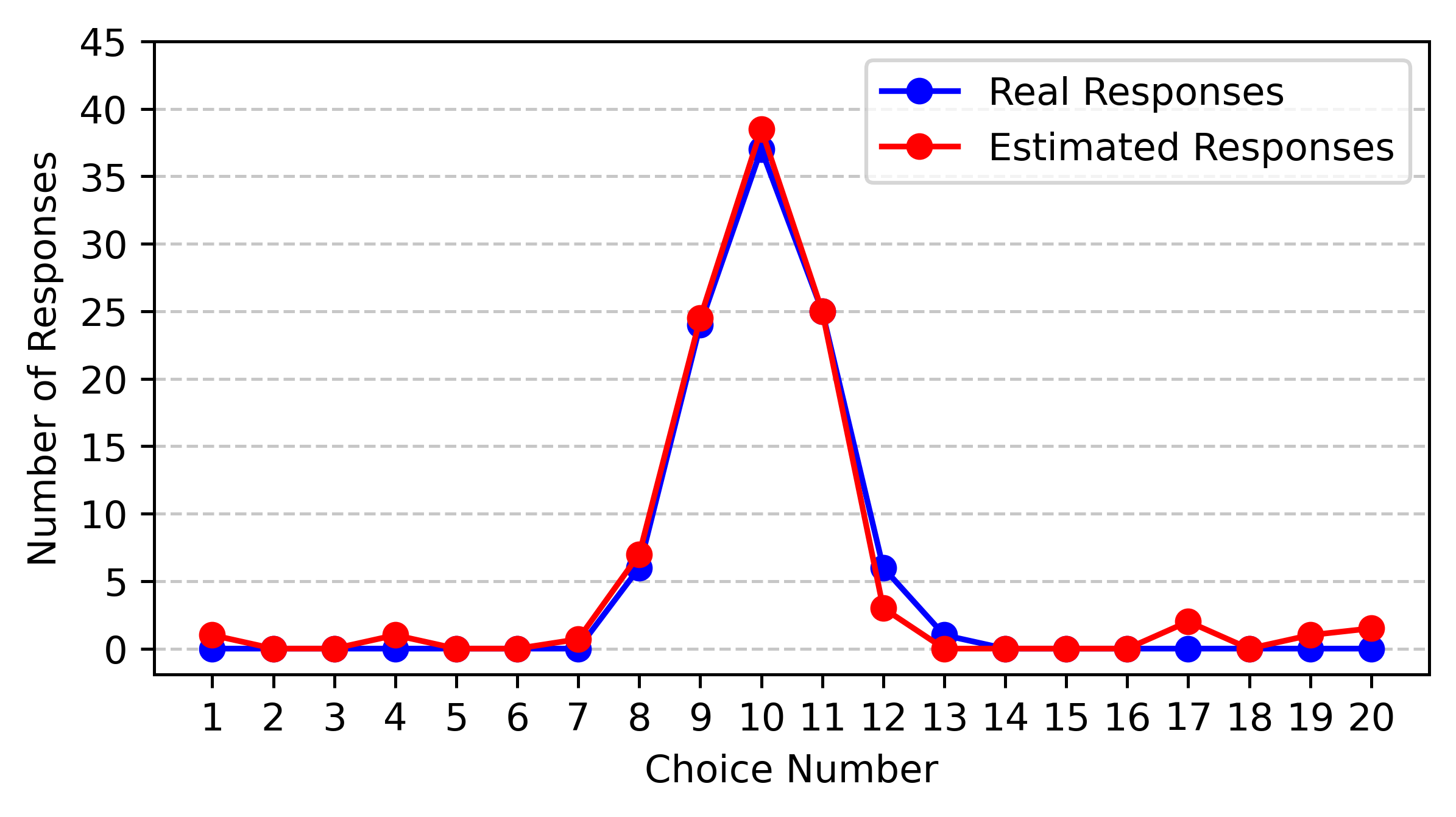}
        \caption{1000 Drones}
        \label{fig:subfig2}
    \end{subfigure}
        \hfill
    \begin{subfigure}[b]{0.24\textwidth}
        \centering
        \includegraphics[width=\textwidth]{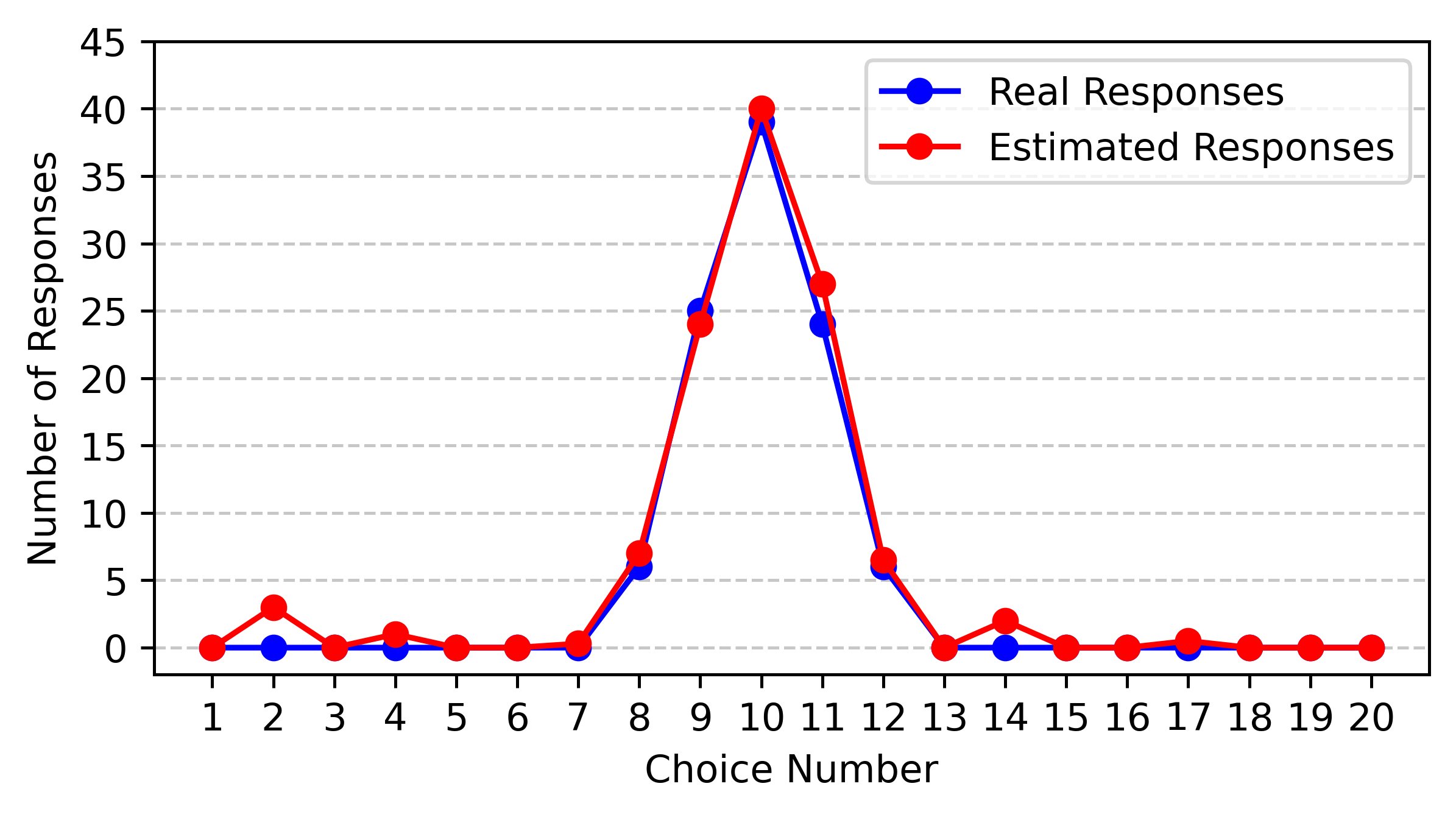}
        \caption{5000 Drones}
        \label{fig:subfig3}
    \end{subfigure}
    \hfill
    \begin{subfigure}[b]{0.24\textwidth}
        \centering
        \includegraphics[width=\textwidth]{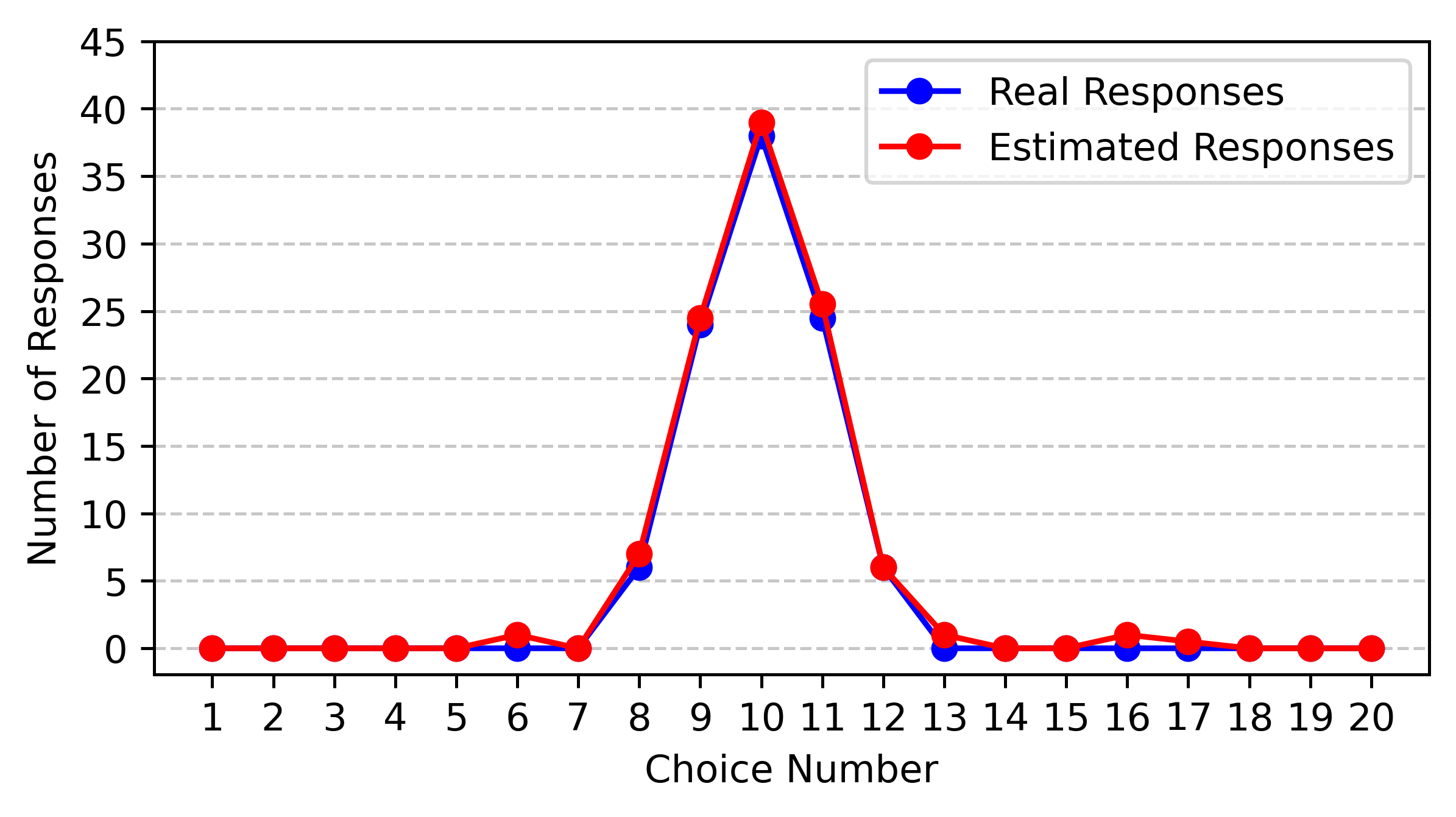}
        \caption{10000 Drones}
        \label{fig:subfig4}
    \end{subfigure}
    \caption{Impact of local differential privacy on data accuracy with different drone counts. Blue line shows the actual response distribution, red line shows the estimated responses post-privacy implementation.}
    \label{fig:figure3}
\end{figure}

The efficiency of local differential privacy (RAPPOR algorithm) is evaluated based on the number of drone operator responses. The experiment simulates a realistic bushfire data collection scenario with 20 possible choices, varying the number of drone operators (500, 1,000, 5,000, and 10,000). The responses are drawn from a normal distribution (mean=10, standard deviation=2). Figure \ref{fig:figure3} compares actual response distributions (blue line) with noisy estimates post-privacy implementation (red line). The experiment shows that despite the number of operators or distribution variance, the accuracy of the extracted results remains invariant, affirming the robustness of the approach.

\subsection{Blockchain-based Overheads}
A smart contract was deployed on the Goerli Ethereum test network to evaluate the blockchain-based overheads. Table \ref{tab:smart_contract_costs} lists the gas costs for each operation, translated into fiat currency. Costs remain independent of the number of drone operators and choices, ensuring scalability for bushfire management scenarios.

\begin{table}[t]
\centering
\caption{Smart Contract Costs. The fiat costs are based on prices from https://useweb3.xyz/gas as of 2 February 2023.}
\begin{tabular}{|l|c|c|}
\hline
\textbf{Operation} & \textbf{Cost in Gas} & \textbf{Cost in Fiat} \\ \hline
Contract Deployment & 660,809 & \$25.40 \\ \hline
Record \( H(C_R) \) & 74,537 & \$2.87 \\ \hline
Record \( H(F) \) & 63,309 & \$2.43 \\ \hline
Make Deposit & 23,642 & \$0.91 \\ \hline
Reveal \( s_2 \) and Transfer Deposit & 36,269 & \$1.39 \\ \hline
\end{tabular}
\label{tab:smart_contract_costs}
\end{table}

\subsection{Dispute Resolution}
The smart contract provides an immutable log essential for dispute resolution:
\begin{enumerate}
    \item \( H(C_R) \) is recorded on the blockchain before the response is transmitted. If \( C_R \) is unreadable, the hash is used to verify the response’s integrity.
    \item Filter disputes are resolved via \( H(F) \) recorded by system operators, enabling users to dispute exclusion or unfair compensation based on this publicly verifiable hash.
\end{enumerate}
Blockchain ensures transparency and accountability, enhancing trust in the crowdsourced drone services.

\subsection{Privacy Properties of RAPPOR}
We evaluate RAPPOR's privacy by considering a query \( Q = \{c_1, c_2, \ldots, c_n\} \) with uniformly distributed responses. The attacker’s chance of guessing a correct response \( C \) is \( P_{\text{guess}} = \frac{1}{n} \). With access to the randomized response \( R \), the probability \( P(r_i = 1) \) is:

\begin{equation}
    P(r_i = 1) = \frac{1}{n} \cdot 0.5 + 0.25 \label{eq:prob_ri}
\end{equation}

\begin{figure}[t]
    \centering
    \begin{minipage}[b]{0.31\textwidth}
        \centering
        \includegraphics[width=\textwidth]{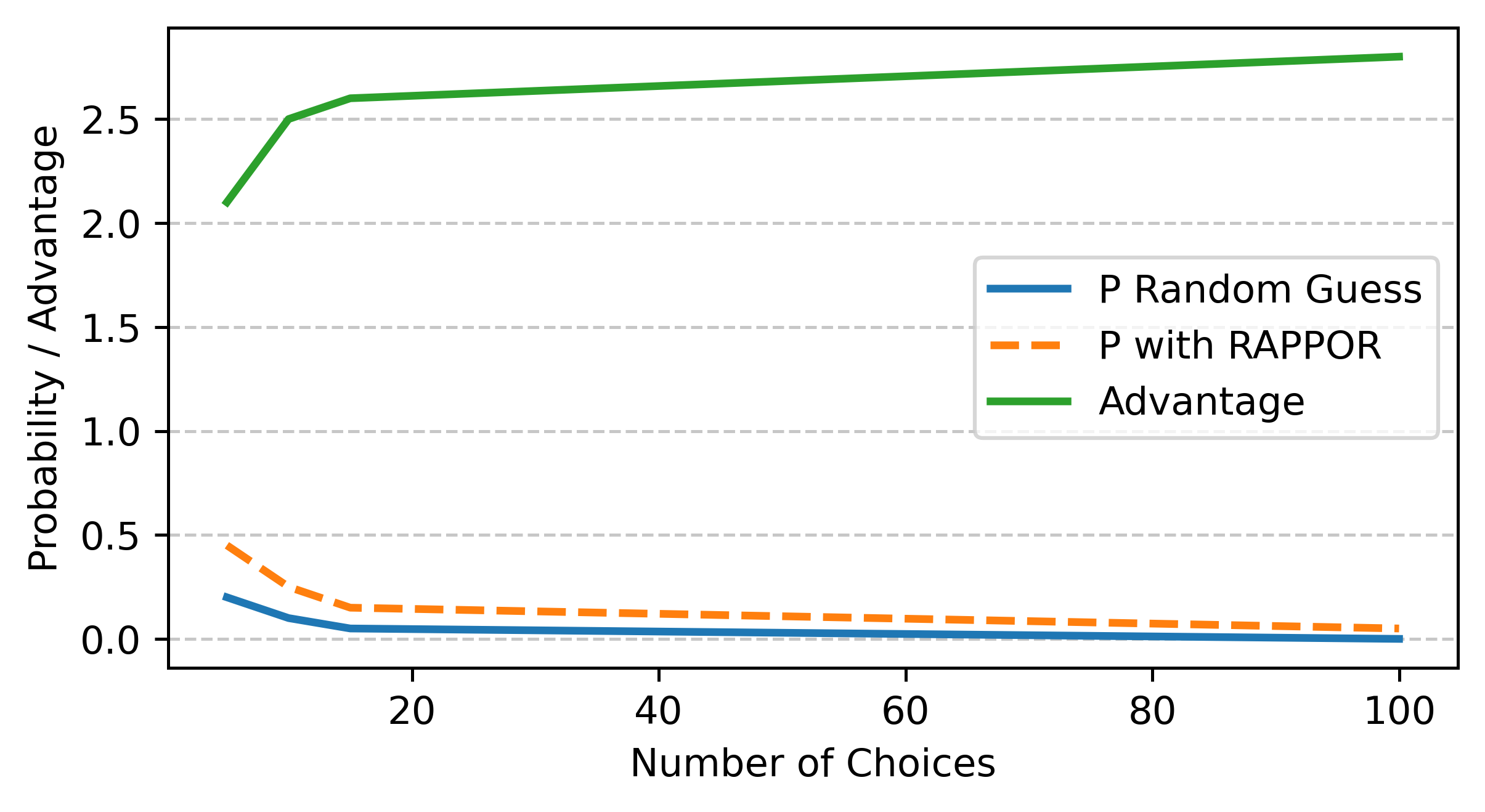}
        \caption{Probabilities and attacker advantage in RAPPOR.}
        \label{fig:figure4}
    \end{minipage}
    \hfill
    \begin{minipage}[b]{0.31\textwidth}
        \centering
        \includegraphics[width=\textwidth]{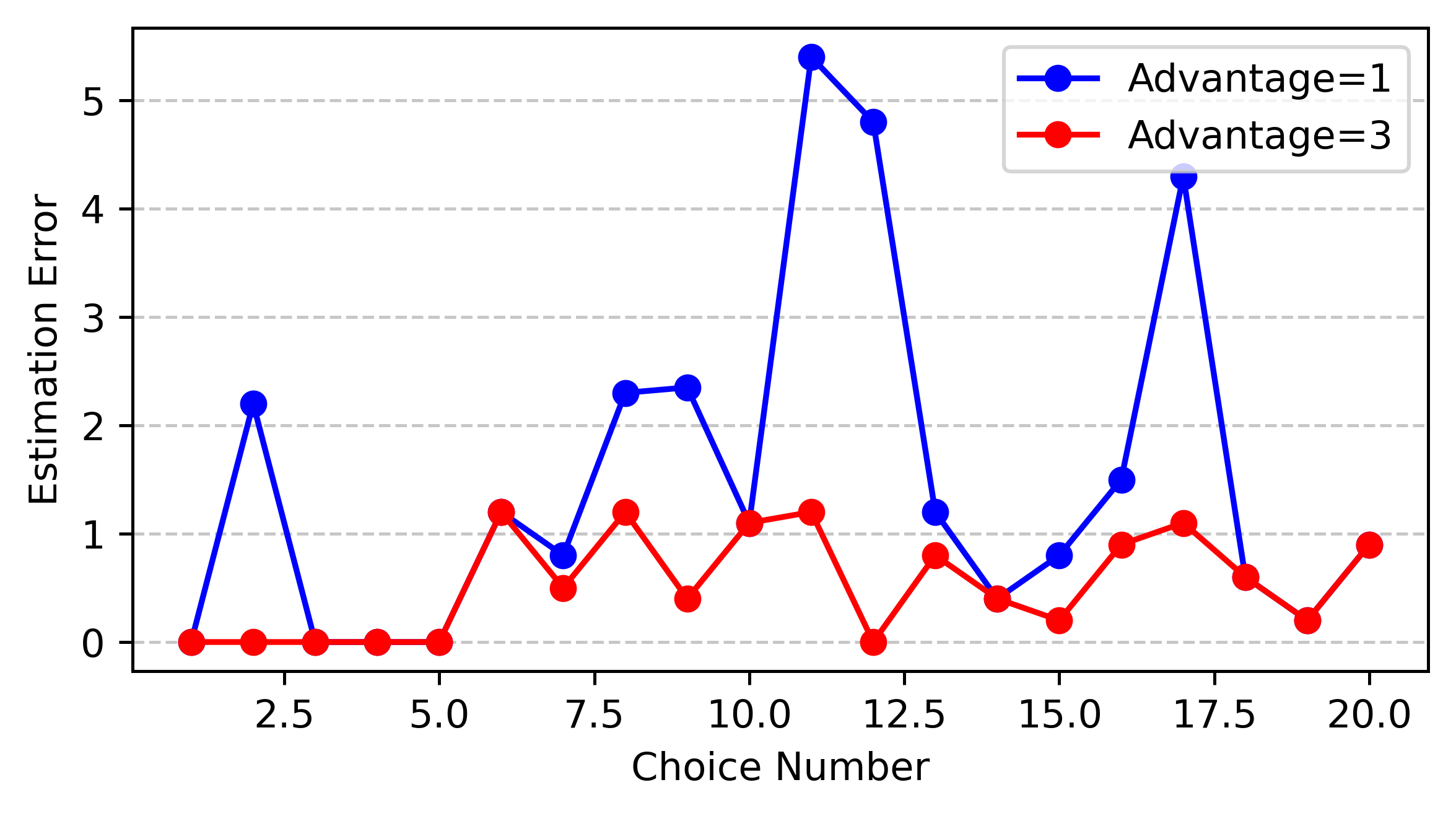}
        \caption{Comparison of response distributions using different coins in RAPPOR.}
        \label{fig:figure5}
    \end{minipage}
    \hfill
    \begin{minipage}[b]{0.31\textwidth}
        \centering
        \includegraphics[width=\textwidth]{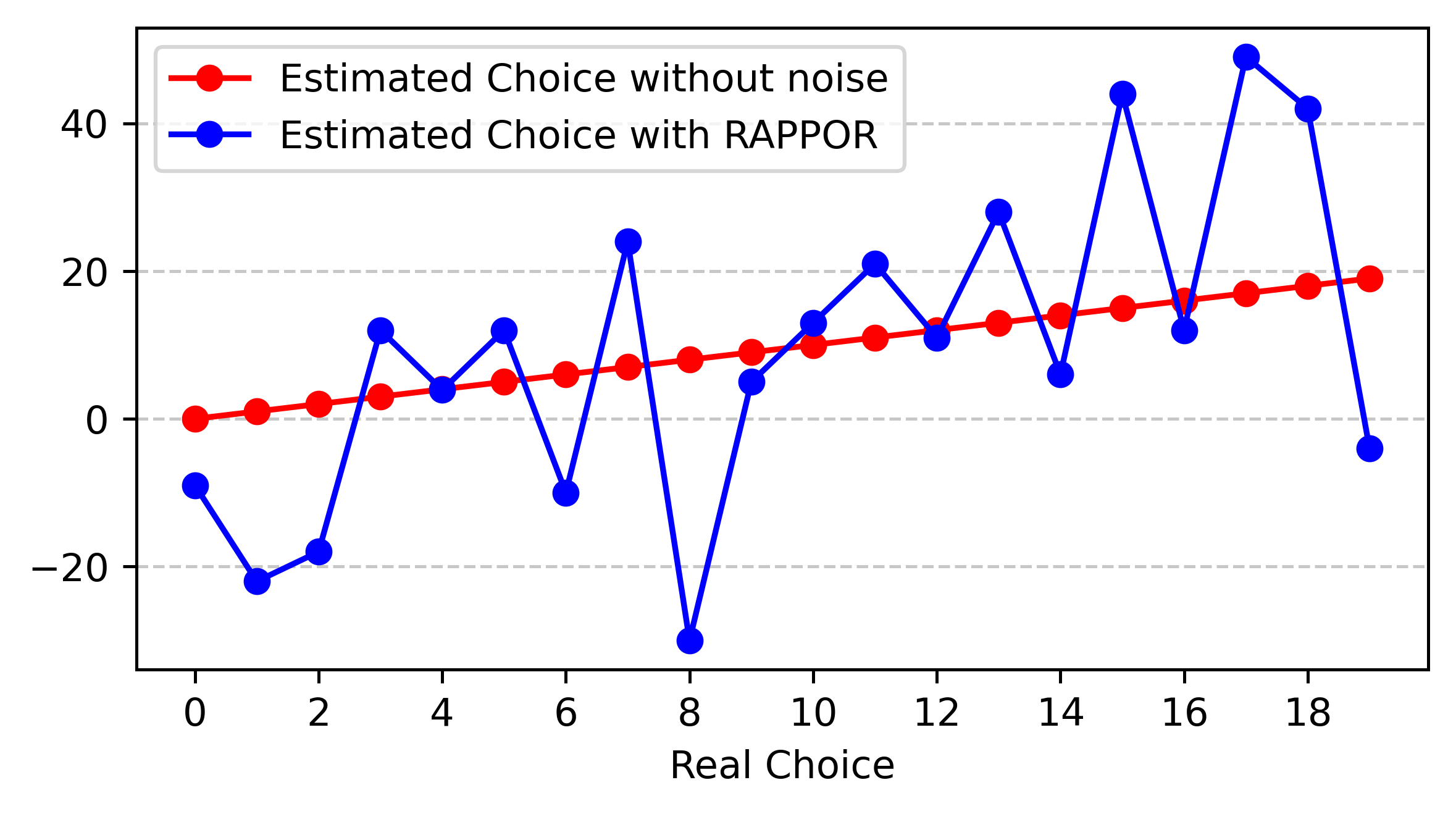}
        \caption{Comparison of the attacker's guessed value (red dots) with and without RAPPOR (blue dots).}
        \label{fig:figure6}
    \end{minipage}
\end{figure}

The attacker’s advantage, \( \text{Adv} = \frac{P_{\text{RAPPOR}}}{P_{\text{guess}}} \), approaches 3 as \( n \) increases (Figure \ref{fig:figure4}). A non-fair coin reducing truth-telling to 20\% lowers the advantage to 1.5 but decreases result accuracy (Figure \ref{fig:figure5}).

\subsection{Privacy-Accuracy Trade-off}
A key threat model involves an attacker observing statistic changes as new data samples are added. Two scenarios are tested:
\begin{itemize}
    \item \textbf{No Noise Addition:} Responses are committed directly without noise, providing accurate statistics.
    \item \textbf{With RAPPOR:} Noise introduced via RAPPOR provides privacy but sacrifices accuracy.
\end{itemize}

Figure \ref{fig:figure6} shows that in the absence of noise, attackers (red dots) can accurately predict responses based on changing averages. In contrast, noise from RAPPOR (blue dots) obfuscates the response, significantly reducing the attacker’s ability to infer individual choices.

\section{Conclusion} \label{conclusion}

In this study, we proposed a privacy-preserving marketplace for crowdsourced drone services in bushfire management. Bushfire management authorities can obtain noisy data from drone operators, essential for deriving meaningful statistics for bushfire detection. We implemented local differential privacy to protect drone operators’ data from all system entities, including system operators, who filter data without accessing raw inputs. Additionally, a blockchain-based infrastructure ensures a fair trade mechanism and immutable records. The system efficiently accommodates multiple stakeholders while maintaining minimal blockchain overhead, independent of the number of drone operators involved.

%A challenge we identified is the potential for dishonest responses from drone operators, a risk inherent in systems employing local differential privacy. To address this, future work will explore the use of Trusted Platform Modules (TPM) for executing drone operator-specific functionalities, enhancing the integrity and reliability of the data. Additionally, while our design minimizes interactions with the blockchain and maintains a fixed cost regardless of stakeholder count, the overhead, complexity, and monetary aspects of employing Ethereum smart contracts are significant considerations. Future research will therefore investigate alternative blockchain models, particularly focusing on private or consortium-based systems, to optimize efficiency and cost-effectiveness in large-scale, real-world applications of our framework for bushfire management.

\bibliographystyle{splncs04}
\bibliography{ref}

%\printbibliography

\end{document}